# Visualizing the emergence of the pseudogap state, and the evolution to superconductivity, in a lightly hole-doped Mott insulator

Y. Kohsaka[1], T. Hanaguri[2], M. Azuma[3], M. Takano[4], J. C. Davis[5,6,7,8], and H. Takagi[1,2,9]

[1] *Inorganic Complex Electron Systems Research Team, RIKEN Advanced Science Institute, Wako, Saitama 351-0198, Japan.*

[2] *Magnetic Materials Laboratory, RIKEN Advanced Science Institute, Wako, Saitama 351-0198, Japan*

[3] *Materials and Structures Lab., Tokyo Institute of Technology, Yokohama, Kanagawa 226-8503, Japan*

[4] *Institute for Integrated Cell-Material Sciences, Kyoto University, Sakyo-ku, Kyoto 606-8501, Japan*

[5] *LASSP, Department of Physics, Cornell University, Ithaca, NY 14853, USA*

[6] *CMPMS Department, Brookhaven National Laboratory, Upton, NY 11973, USA.*

[7] *School of Physics and Astronomy, University of St. Andrews, St. Andrews, Fife KY16 9SS, UK.*

[8] *Kavli Institute at Cornell for Nanoscale Science, Cornell University, Ithaca, NY 14853, USA.*

[9] *Department of Physics, University of Tokyo, Hongo, Tokyo 113-0033, Japan*

**1** **Superconductivity emerges from the cuprate antiferromagnetic Mott state with hole doping. The resulting electronic structure[1] is not understood, although changes in the state of oxygen atoms appear paramount[2-5]. Hole doping first destroys the Mott state yielding a weak insulator[6,7] where electrons localize only at low temperatures without a full energy gap. At higher doping, the 'pseudogap', a weakly conducting state with an anisotropic energy gap and intra-unit-cell breaking of 90º-rotational ($C_{4v}$) symmetry appears[3,4,8-10]. However, a direct visualization of the emergence of these phenomena with increasing hole density has never been achieved. Here we report atomic-scale imaging of electronic structure evolution from the weak-insulator through the emergence of the pseudogap to the superconducting state in $Ca_{2-x}Na_xCuO_2Cl_2$. The spectral signature of the pseudogap emerges at lowest doping from a weakly insulating but $C_{4v}$-symmetric matrix exhibiting a distinct spectral shape. At slightly higher hole-density, nanoscale regions exhibiting pseudogap spectra and 180º-rotational ($C_{2v}$) symmetry form unidirectional clusters within the $C_{4v}$-symmetric matrix. Thus, hole-doping proceeds by the appearance of nanoscale clusters of localized holes** **within which the broken-symmetry pseudogap state is stabilized. A fundamentally two-component electronic structure[11] then exists in $Ca_{2-x}Na_xCuO_2Cl_2$ until the $C_{2v}$-symmetric clusters touch at higher doping, and the long-range superconductivity appears.**

2  To visualize at the atomic scale how the pseudogap and superconducting states are formed sequentially from the weak insulator state, we performed spectroscopic imaging scanning tunneling microscopy (SI-STM) studies on $Ca_{2-x}Na_xCuO_2Cl_2$ ($0.06 \leq x \leq 0.12$) (see also the Methods sections). The crystal structure is simple tetragonal (I4/mmm) and thereby advantageous because the $CuO_2$ planes are unbuckled and free from the orthorhombic distortion. More importantly $Ca_2CuO_2Cl_2$ can be doped from the Mott insulator to the superconductor by introduction of Na atoms. Figs. 1c and 1d show differential conductance $dI/dV(r,E=eV)$ images measured using SI-STM of bulk-insulating $x = 0.06$ and $x = 0.08$ samples taken in the field of views of the topographic images in Figs. 1a and 1b. The wavy arcs bright in color in Figs. 1c and 1d have never been observed in superconducting samples ($x > 0.08$) but appear only in such quasi-insulating samples ($x \leq 0.08$). They are created by spectral peaks in $dI/dV(r,E=eV)$ whose energy is dependent on location (Fig. 1f ). Consequently, the wavy arcs shrink with increasing bias voltages and finally disappear. This behavior, due to tip-induced impurity charging[12-14], is characteristic of poor electronic screening in a weakly insulating state.

3  A wide variety of spectral shapes originating from electric heterogeneity were found in these samples. A typical example of the spectra is, as spectrum #1 in Fig. 1e, the V-shaped pseudogap (~0.2 eV) spectrum with small dip (~20 meV) near the Fermi energy. This is indistinguishable from those found in strongly underdoped cuprate superconductors[3], and establishes that the pseudogap state appears locally at the nanoscale within the weak-insulator. Besides the V-shaped pseudogap spectra in some areas, we find a new class of spectra that is predominant elsewhere in the insulating samples. As for example spectrum #2 in Fig. 1e, such spectra are extremely asymmetric about the Fermi energy, U-shaped (concave in minus a few hundreds of mV), and exhibit no clear pseudogap. The growing asymmetry is strongly indicative of approaching the Mott insulating state[15,16] while the non-zero conductance in the unoccupied state is distinct from the Mott insulating state[17]. The approach for spectroscopic examination of the emergence of the pseudogap from the weak-insulator is therefore transformation from the U-shaped insulating spectra to the V-shaped pseudogap spectra as a function of location and doping.

4    Fig. 2a represents the transformation between these two types of spectra. The V-shaped pseudogap becomes larger and broader, and eventually is smoothly connected to the U-shaped insulating spectra. To quantify this variation, we focus on positive biases where the edge of the pseudogap is clear. We fit the following function to each spectrum[18],

$$f(E) = c_0 \, \text{Re}\left[\frac{E + i\Gamma(E)}{\sqrt{(E + i\Gamma(E))^2 - \Delta^2}}\right] + c_1 E + c_2 \qquad (1)$$

where $E$ is the energy, $\Gamma$ is the broadening term, $\Delta$ is the energy gap, and $c_i$ ($i = 0, 1, 2$) are fitting constants. Use of Eqn. 1 is merely for accurate quantitative parameterization of the gap maximum and does not imply any particular electronic state. We use $\Gamma(E) = \alpha E$ as Ref. 18 ($\alpha$ is a proportional constant) but momentum-independent $\Delta$ for simplicity of fitting procedures (See also Supplementary Information). The results of fits are shown by the black curves superimposed on the experimental data in Fig. 2a. The excellent agreement between the data and the fits ensures the precision of extracted parameters. Moreover, the extracted $\Delta$ and $\alpha$ demonstrates that this fitting works to parameterize the variation of spectra through magnitude ($\Delta$) and broadening in energy of the states at pseudogap energy.

5    On the basis of the successful fits, we plot spatial variation and doping evolution of $\Delta$ and $\alpha$ in Figs. 2b – 2e. The impurity charging clearly appears as the wavy arcs red and orange in $\alpha$-maps, preventing us from extracting $\alpha$ associated with the pseudogap where the wavy arcs appear. Nevertheless, the $\Delta$- and $\alpha$-maps exhibit large cross-correlations (correlation coefficients are 0.55 for $x = 0.06$ and 0.59 for $x = 0.08$), corroborating that the pseudogap spectrum gets broader with its increasing magnitude as holes are removed. Purple and blue areas are predominant in the $x = 0.06$ sample (Fig. 2b). Spectra found in these areas are the U-shaped ones of the weak-insulator as shown in Fig. 2a in the same color. Orange and yellow areas in the $\Delta$-maps show clear pseudogap as shown in Fig. 2a. Thus we find that, for bulk-insulating samples, the nanoscale pseudogap regions are embedded within the matrices of the weak-insulator (Fig. 2b). As more holes are introduced, the pseudogap areas eventually become predominant and mutually connected near the insulator-superconductor critical doping (Fig. 2c). They eventually overwhelm the insulating areas of the underdoped superconducting sample (Fig. 3b).

6   The high spatial resolution of this new pseudogap analysis reveals that the spatial variation of $\Delta$ itself, is locally ordered rather than randomly dispersed. The atomically resolved $\Delta$-maps (Figs. 3a-3d) clearly show that the spatial arrangements of the pseudogap energy $\Delta(r)$ is typically elongated in a Cu-O bond direction and forms bond-like objects on Cu-O-Cu complexes. Such bond-like objects then align to each other in the direction normal to the bonds and organize themselves into nm-scale unidirectional clusters. Given the excellent agreement of the fit, these ordered structures in the $\Delta$-maps indicate that the spectral shape is spatially modulated over quite a wide energy range. Therefore, spatial variation of the local density-of-states (LDOS) extracted appropriately is expected to have similar spatial structures.

7   To confirm this, we calculate the $R$-maps, $R(\bm{r},E) \equiv I(\bm{r},+E) / I(\bm{r},-E) = \int_0^E N(\bm{r},\varepsilon)\mathrm{d}\varepsilon \big/ \int_{-E}^0 N(\bm{r},\varepsilon)\mathrm{d}\varepsilon$, where $I$ and $N$ are the tunneling current and the LDOS, respectively, at the location $\bm{r}$ and the energy $E$ (Ref. 3). By choosing an appropriate energy window of the integration, we can map the LDOS variation around the pseudogap energy onto a single image of the $R$-map without influence of the intense electronic heterogeneity[19]. As shown in Figs. 3e and 3f, the $R$-maps obviously show the same structures as those of the $\Delta$-maps in both atomic and nm scale. We therefore conclude that the pseudogap state breaks translational and rotational symmetry of the lattice in the form of the unidirectional clusters consisting of the bond-like objects on Cu-O-Cu complexes.

8   To analyze the broken symmetry observed in the Δ- and the $R$-maps, we focus on local symmetry of the maps. To separate $C_{2v}$ symmetry of the bond-like objects from $C_{4v}$ symmetry of the lattice, we measure the following two components about each unit cell, $Q_{xx}(f;\mathbf{r}_{Cu}) = \sum_{i=1}^{4}(-1)^{i}f(\mathbf{r}_{O(i)})$ and $Q_{xy}(f;\mathbf{r}_{Cu}) = \sum_{i=1}^{4}(-1)^{i}f(\mathbf{r}_{O'(i)})$, where $f$ is a two dimensional image, $\mathbf{r}_{O(i)}$ is the locations of four oxygen atoms surrounding a copper atom at the location $\mathbf{r}_{Cu}$ as illustrated in the inset of Fig. 4h. $\mathbf{r}_{O'(i)}$ is the location given by 45º rotation of $\mathbf{r}_{O(i)}$ about the central copper. Either $Q_{xx}$ or $Q_{xy}$ (or both) is non-zero for $C_{2v}$ patterns and both of them are zero for $C_{4v}$ patterns (See also Supplementary Information).

9   The $Q_{xx}$ and $Q_{xy}$ about the Δ- and $R$-maps shown in Fig. 4 quantitatively visualize how those maps breaks the lattice symmetry and forms short-range order. The bond-like objects and the nm-scale unidirectional clusters are expressed as bars and arrays of bars, as exemplified by the rectangles in Figs. 3 and 4. These bars and arrays are closely arranged in $Q_{xx}$ rather than $Q_{xy}$, evidently showing the $C_{2v}$ symmetry breaking occurs in the direction not diagonal but parallel to the Cu-O bond. Meanwhile, as shown by the ovals in Figs. 4a-4d, areas without clear bars in both $Q_{xx}$ and $Q_{xy}$ are found in $x = 0.08$, indicating the symmetry of patterns is $C_{4v}$. Given that U-shaped insulating spectra are found in large-Δ areas while V-shaped pseudogap spectra in small-Δ areas, we find that larger electron-hole asymmetry (smaller $R$) and U-shaped insulating spectra are found in the $C_{4v}$ areas while smaller electron-hole asymmetry (larger $R$) and clear V-shaped pseudogap spectra are found in the $C_{2v}$ areas by comparing Figs. 3 and 4. This further supports that the pseudogap state breaks the symmetry of lattice. Large correlation between $Q_{xx}(\Delta)$ and $Q_{xx}(R)$, -0.45 for Figs. 4a and 4c and -0.60 for Figs. 4e and 4g, quantitatively confirms similarity between the Δ- and $R$-maps about the local broken symmetry.

10   Our results reveal new perspectives on the genesis of the cuprate pseudogap state[20]. Firstly, the pseudogap regions emerge at lowest dopings as nm-scale clusters which are embedded in the $C_4$-symmetric matrices of a weak-insulator. This implies that short-range localization of the doped holes and strong interaction between them is the first step in formation of the pseudogap state out of the Mott insulator. Second, we discovered that the pseudogap energy scale itself $\Delta(\boldsymbol{r})$ exhibits strong breaking of $C_4$-symmetry down to the $C_{2v}$ symmetry, at the unit-cell scale within these clusters. Third, the spatial boundaries between $C_{2v}$ (pseudogap) and $C_{4v}$ (insulating) areas can be atomically sharp as highlighted by neighboring the ovals and the rectangle in Figs 3a and 3e. This indicates that while the pseudogap state is distinctive electronic phase, it can exist as short-range order[11]. Fourth, the width of $C_{2v}$-clusters is constantly about $4a_0$ ($a_0$: distance between nearest copper atoms) independently of doping of $0.06 \leq x \leq 0.12$ (see also Supplementary Information). Moreover, the $C_{2v}$-clusters are aligned with the Cu-O bonds in the whole doping range studied, indicating that this form of unit-cell-scale rotational symmetry breaking[3,4] is an elementary characteristic of the pseudogap phase.

11  All of the above observations reveal that hole doping into the $C_4$-symmetric weak-insulator is not a homogenous process in $Ca_{2-x}Na_xCuO_2Cl_2$. Rather it proceeds by localization of doped holes into $C_{2v}$-symmetric nano-clusters embedded in the $C_4$-symmetric insulator so that the bulk materials must exhibit a fundamentally heterogeneous two-component electronic structure[11]. This situation persists until, with increasing dopant density, the $C_{2v}$ pseudogap clusters become interconnected (percolate) at $x = 0.08$, coincident with the appearance of long-range superconductivity. This observation appears to imply a beneficial rather than competing role for the pseudogap phase in the appearance of cuprate high temperature superconductivity.

**Methods**

12  The $Ca_{2-x}Na_xCuO_2Cl_2$ crystals used in this study were grown by the flux method under high pressure of several GPa[21]. The samples of $x = 0.06$, $0.08$, and $0.12$ are insulators, insulator (in the close vicinity of the insulator-superconductor critical doping), and underdoped superconductors of $T_c = 21$ K, respectively[21,22]. The samples were cleaved at 77 K or below and immediately transferred to the heads of microscopes cooled down beforehand. All measurements were done at 4.6 K or below with tungsten tips. The tips were prepared by field emission on gold or field evaporation with field ion microscopes. Differential conductance was measured by the standard lock-in technique with modulation voltages of 1 – 5 mV. The bias voltages are applied to the sample.

**Acknowledgements**

We gratefully acknowledge discussions with Michael Lawler and Eun-Ah Kim. Studies at Brookhaven/Cornell are supported by the Center for Emergent Superconductivity, an Energy Frontier Research Center, under DE-2009-BNL-PM015, and studies at RIKEN by JSPS KAKENHI (19840052, 20244060).

**Author contributions**

Y. K, M. A, and M. T grew single crystals. Y. K. and T. H. performed STM measurements. Y. K. analyzed data. Y. K. and J. C. D. wrote the manuscript. J. C. D and H. T supervised the project.

**Competing financial interests**

The authors declare no competing financial interests.

**Figure captions**

**Figure. 1 Spectral features unique to the insulating samples.**

**a**, **b**, 20 × 20 $nm^2$ square constant-current topographic images of $x = 0.06$ and $x$ 0.08, respectively. Scanning parameters are 0.1 nA at -0.4 V for **a** and 0.1 nA at -0.3 V for **b**. The markers with numbers indicate locations where spectra shown **e** and **f** were taken.

**c**, **d**, Differential conductance maps taken at -0.28 V and -0.22 V in the same of views as **a** and **b**, respectively. Scanning parameters are 0.2 nA at -0.4 V for **c**, and 0.15 nA at -0.3 V for **d**. The wavy arcs in bright color caused by the tip-induced impurity charging guarantee that these surfaces are insulating.

**e**, **f**, Examples of differential conductance spectra taken in the insulating samples. Numbers denote locations where these spectra were taken shown in **a** and **b**. Peaks found in the spectra of **f** are caused by the tip-induced impurity charging. The setup conditions of spectra in **e** and **f** are 0.2 nA at -0.4 V and 0.15 nA at -0.3 V, respectively.

**Figure. 2 Spatial variations and doping evolution of the pseudogap.**

**a**, Examples of differential conductance spectra taken various locations of multiple samples of $0.06 \leq x \leq 0.12$. Each spectrum is shifted vertically for clarity and color-coded based on the values of $\Delta$ with the color scale same as used in **b** and **c**. The horizontal markers indicate zero of each curve. The black curves superimposed on the spectra are the results of fits described in the text. The vertical markers denote extracted $\Delta$ while the numbers denote $\alpha$.

**b**, **c**, **d**, **e**, $20 \times 20$ nm$^2$ square $\Delta$-maps and $\alpha$-maps of $x = 0.06$ and $x = 0.08$. The original $256 \times 256$ spectra were taken in the same fields of view as Figs. 1a and 1b, respectively. Note that color scales are common to **b** and **c**, and **d** and **e**. The box of white dotted line in **c** shows the area of Fig. 3a.

**Figure. 3 Atomic and nano-scale short range order of the pseudogap state found in $\Delta$-maps and $R$-maps.**

**a**, **b**, $12 \times 12$ nm$^2$ square $\Delta$-maps of $x = 0.08$ and $x = 0.12$, respectively. **a** was taken in the area shown by the white-dotted box in Fig. 2c. **b** was taken in the same area of the same sample used in Ref. 3. The color scale is the same as used in Figs. 2b and 2c. The black arrows indicate directions of Cu-O bonds. The solid lines in **b** are trajectories along which spectra shown in **c** were taken.

**c**, Differential conductance spectra of $Ca_{1.88}Na_{0.12}CuO_2Cl_2$ taken along the line 1 and 2 shown in **b**, demonstrating $\Delta$ is actually modulated at the atomic scale. Each observed spectrum black in color are shifted vertically for clarity. Results of the fit, blue and orange in color, are superimposed on the observed spectra. The markers denote extracted $\Delta$.

**d**, $\Delta$ extracted from fits to the spectra shown in **c**. The inset depicts location of the trajectories where the spectra shown in c were taken, relative to the $CuO_2$ plane. $\Delta$ is modulated along the trajectories in the same manner at the atomic scale, leading to the formation of bond-like objects on Cu-O-Cu complexes.

**e**, **f**, 12 × 12 nm$^2$ square *R*-maps taken in the same areas as **a** and **b**, respectively. The integration voltages are 252 meV for **e** and 150 meV for **f**. The white arrows indicate directions of Cu-O bonds.

The ovals and rectangles of broken lines are guides to the eye indicating an area with no clear pseudogap, and clear pseudogap and the nm-scale unidirectional clusters consisting of the bond-like objects, respectively.

**Figure. 4 Quantitative analysis of local symmetry corroborating that the pseudogap state accompanies local C$_2$ symmetry.**

The values of $Q_{xx}$ and $Q_{xy}$ at each copper site indicated by the gray dots are expressed by the short bars. The length of bars denotes magnitude of $Q_{xx}$ and $Q_{xy}$. The color of bars carries the same information as the length does for clarity. The direction of each bar corresponds to sign of $Q_{xx}$ and $Q_{xy}$. The panels of left (**a**, **c**, **e**, **g**) and right (**b**, **d**, **f**, **h**) columns are $Q_{xx}$ and $Q_{xy}$, respectively. The upper (**a**-**d**) and lower (**e**-**h**) four panels are computed for $x = 0.08$ and $x = 0.12$, respectively. The ovals in **a** - **d** and the rectangles in **a**, **c**, **e**, and **g** are drawn at the same location as those in Fig. 3. The first and third rows (**a**, **b**, **e**, **f**) and second and fourth rows (**c**, **d**, **g**, **h**) are computed for the $\Delta$- and *R*-maps, respectively. The original $\Delta$- and *R*-maps are shown in Fig. 3. The inset of **h** depicts that locations of $r_{Cu}$, $r_{O(i)}$, and $r_{O'(i)}$ used in the definition of $Q_{xx}$ and $Q_{xy}$ described in the text.

**Fig. 1**

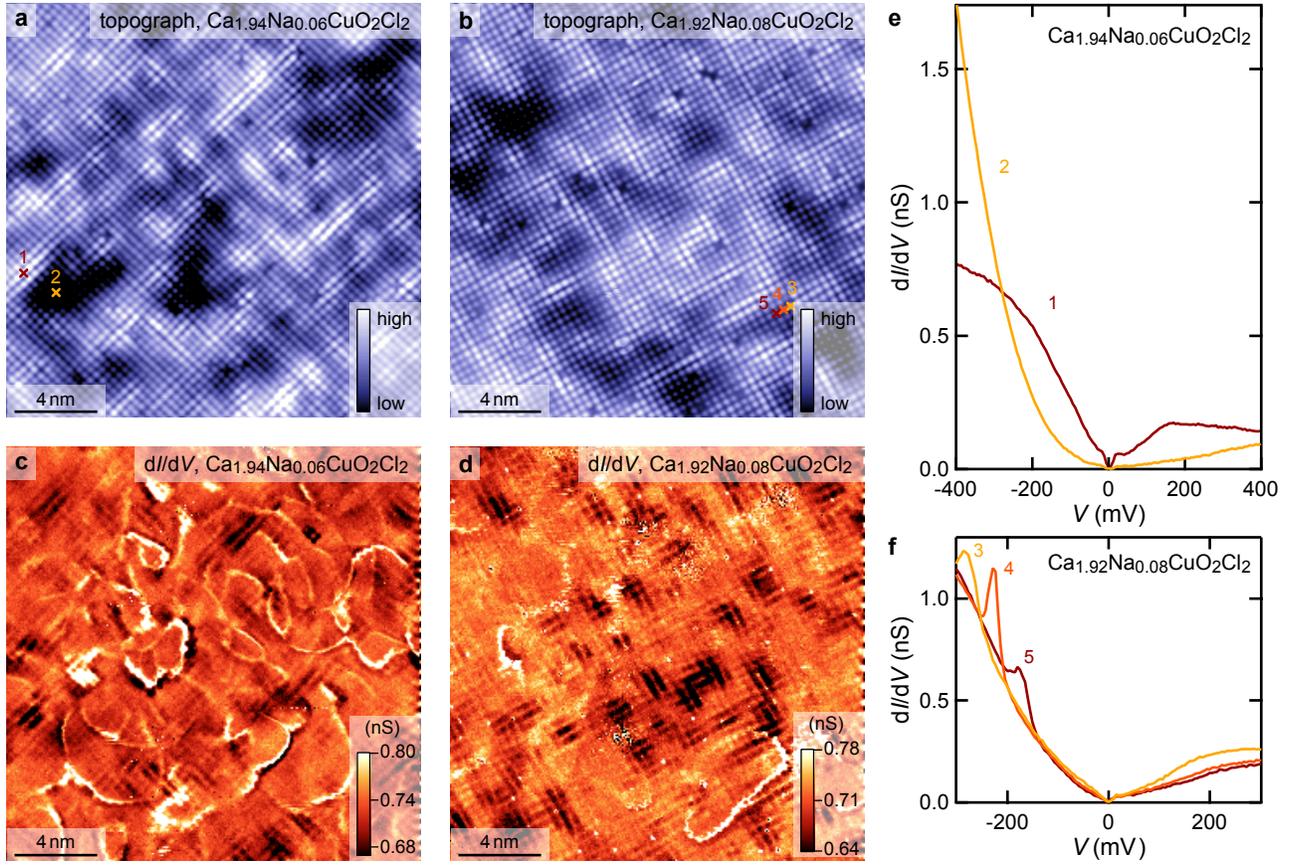

Fig. 2

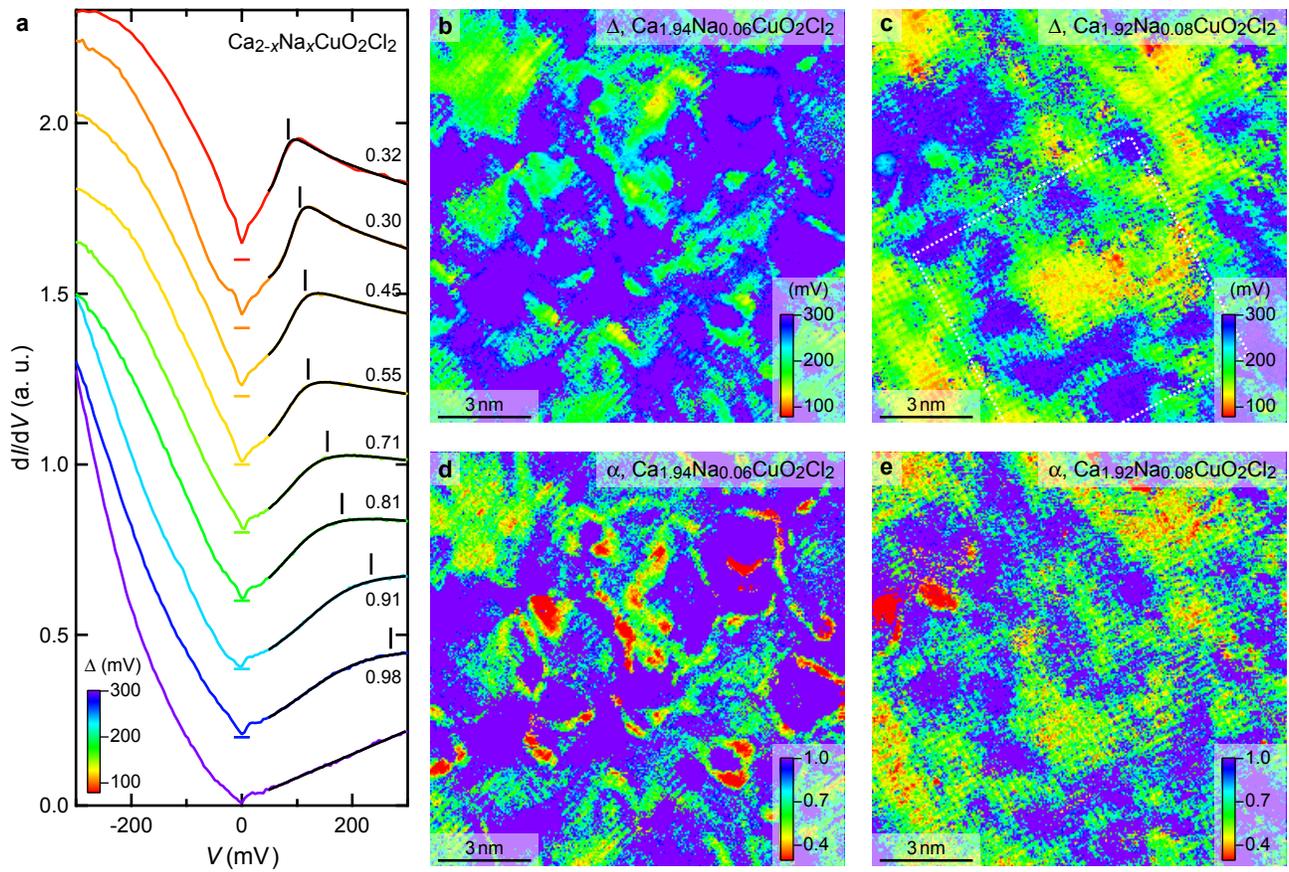

**Fig. 3**

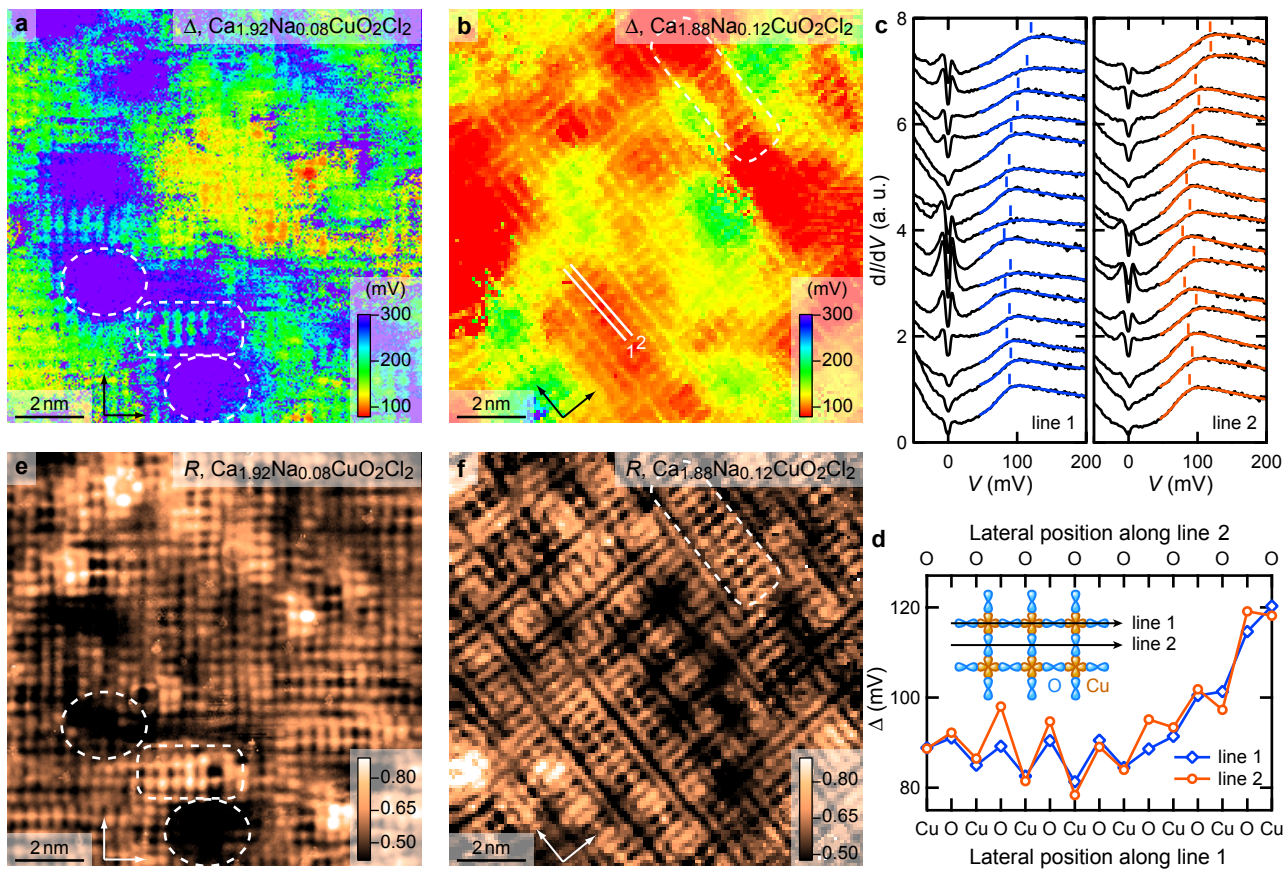

**Fig. 4**

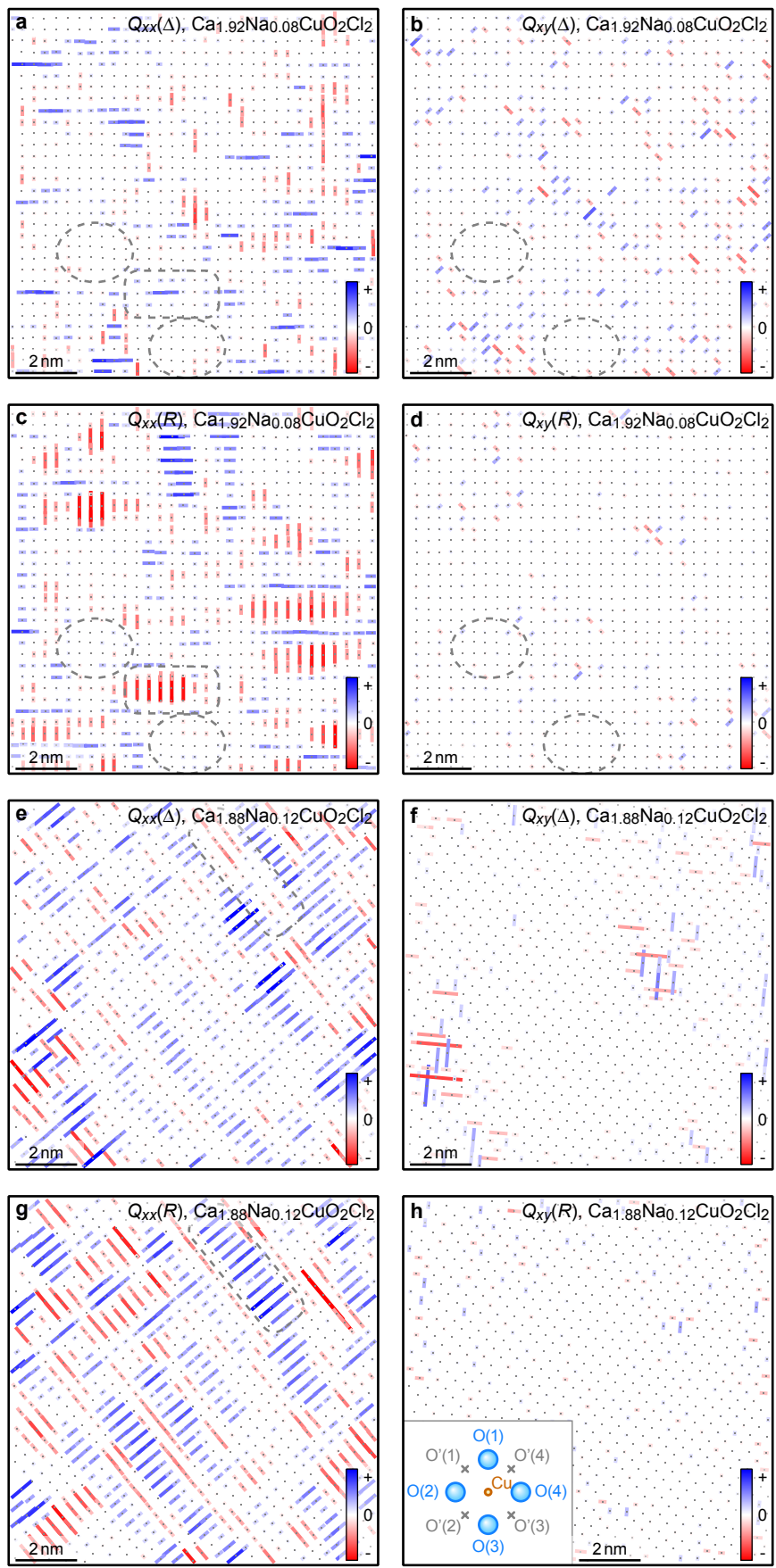